\input{aipcheck}

\documentclass[final]{aipproc}   
\usepackage{graphicx}
\usepackage{amssymb}
\usepackage{epstopdf}      

\layoutstyle{8x11double}

\begin{document}

\title{VERITAS Observations of MGRO J1908+06/HESS J1908+063}


\author{John E Ward$^*$ for the VERITAS Collaboration$^\dagger$}{
  address={$^*$
School of Physics, University College Dublin, Belfield, Dublin, Ireland} 
  ,altaddress={$^\dagger$For a full list of authors, see: http://veritas.sao.arizona.edu}
}



\begin{abstract}
The unidentified TeV gamma-ray source MGRO J1908+06/HESS J1908+063 was observed with the VERITAS Imaging Atmospheric Cherenkov Array during
October 2007 and May-June 2008. This extended source is located on the galactic plane at a galactic longitude of 40.45 degrees and has a hard TeV spectrum
with an index of $\Gamma$$\approx$2.08 \cite{hess_milagro}. The Very High Energy (VHE) gamma-ray flux was measured by H.E.S.S. out to energies greater than 30 TeV which along with its unidentified nature makes it an
interesting hard-spectrum extended source for study. We confirm the detection of VHE gamma-ray emission from this source using VERITAS.

\end{abstract}

\maketitle


\section{Introduction}

The Milagro collaboration reported the detections of several $\gamma$-ray sources in 2007 which
included the unidentified source MGRO J1908+06 \cite{milagro_survey07short} at an 8.3$\sigma$ (pre-trials)
level. Milagro report a source extension up to 2.6$^\circ$ with 90$\%$ confidence.

The H.E.S.S. experiment's extended Galactic plane survey \cite{hess_survey07} has confirmed
this source with a post-trials estimate of 5.7$\sigma$. H.E.S.S. reported a flux
above 300 GeV of 14\% of the Crab Nebula flux, a best-fit position for the emission region of l=40.45$\pm$0.06$^\circ$$_{stat}$$\pm$0.06$^\circ$$_{sys}$ (19$^h$08$^m$04.39$^s$) and b=-0.8$\pm$0.05$^\circ$$_{stat}$$\pm$0.06$^\circ$$_{sys}$ (06$^\circ$19$'$09.1$''$) along with an intrinsic source extension of 
$\sigma_{src}$=0.21$^\circ$$\pm$0.07$^\circ$$_{stat}$$\pm$0.05$^\circ$$_{sys}$ \cite{hess_milagro}.

The source may be associated with the radio-bright SNR G40.5-0.5 \cite{downes} or possibly the EGRET sources 3EG J1903+063 \cite{1999yCat..21230079H} and GeV J1907+0557 \cite{1997ApJ...488..872L}. However, the possibility of TeV
emission from nearby unresolved sources and/or cosmic-ray interaction with nearby CO molecular clouds cannot be discounted. 
The VERITAS measurements of the source position and size along with a discussion on possible counterpart sources are presented here.

The extended nature of this source, along with its relative strength make it an excellent analysis calibration source for the VERITAS sky survey.


\section{Observations}


VERITAS first observed the quoted H.E.S.S. position for TeV emission in
October and November 2007 with further follow-up observations taken
in May and June 2008. Two
differing modes of observation were used to observe this
source. 
The first observations were made using the concept of a "mini-survey". A grid of pointed array observations with multiple camera off-sets (from the reported H.E.S.S. position) was created to emulate the conditions of the main VERITAS sky survey.

A second set of observations were then undertaken that consisted purely of "Wobble" runs
with offsets of 0.5$^\circ$ and 0.7$^\circ$ from the reported
H.E.S.S. position. 
The average
camera off-set for all the taken data is 0.65$^\circ$. 

All data used in the analysis have passed data-quality cuts (very good weather coupled with no hardware issues) giving a
total of $\sim$1300 minutes of live-time data. A table containing a break-down of the total time into it's constituent modes is shown (Table 1.).
The zenith angles of observations ranged from 25 to 50$^\circ$ with an average zenith of 34$^\circ$.

\begin{table}
\begin{tabular}{lrrrr}
\hline
  & \tablehead{1}{r}{b}{Survey-style\\ pointings}
  & \tablehead{1}{r}{b}{0.5$^\circ$\\ wobble}
  & \tablehead{1}{r}{b}{0.7$^\circ$\\ wobble}
  & \tablehead{1}{r}{b}{Total}\\
\hline
Time(min) & 340 & 300 & 660    & 1300\\
\hline
\end{tabular}
\caption{Breakdown of VERITAS observations}
\label{tab:a}
\end{table}





\section{Analysis and Results}

\subsection{VERITAS analysis}

The VERITAS data analysis steps consist of 

\begin{itemize}

\item Calibration and integration of the flash-ADC traces recorded when the array is triggered by a Cherenkov flash

\item Individual telescope Cherenkov images are then cleaned and parameterized according to Hillas \cite{1985ICRC....3..445H}.

\item Stereoscopic reconstruction of the event impact position and direction, hadronic background rejection and energy estimation are undertaken. 

\item Generation of photon maps.

\item The Ring Background Model analysis \cite{2007A&A...466.1219B} is then used to estimate the hadronic background in the source region. 

\end{itemize}

Several changes were made to the standard VERITAS analysis for this source that are highly motivated by the published H.E.S.S. results (spectral index and extension values). Since both the H.E.S.S. and Milagro instruments detail that this source is not point-like, a larger background ring radius was chosen to avoid contamination of the background estimation. H.E.S.S. also quote a spectral index of $\alpha$ = 2.08, making this a hard-spectral source in the TeV regime. Since the VERITAS observations of this source were on average taken at relatively low elevations, a harder size cut of 850 dc 
($\approxeq$150 p.e) is well motivated (studied with simulations) and is used in this analysis. 

To avoid gamma-ray contamination in the background ring, an exclusion region of 0.6$^\circ$ radius was placed around the reported H.E.S.S. position.


\subsection{Results}

MGRO J1908+06/HESS J1908+063 has been detected by VERITAS at the 4.9$\sigma$ level. This quoted significance value is at the reported H.E.S.S. position. 
Figure 1 shows the smoothed significance map of the region with the 4.9$\sigma$ VERITAS value at the H.E.S.S. position marked with a cross. Figure 2 shows the distribution of significances for the field of view around the source.

\begin{figure}
\includegraphics[height=.3\textheight]{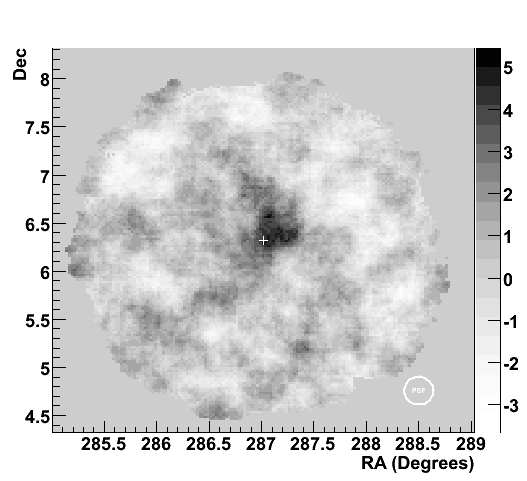}
\caption{Significance Map of the FOV around MGRO J1908/HESS J1908. The cross marks the reported H.E.S.S. location (4.9$\sigma$)}
\end{figure}


\begin{figure}
\includegraphics[height=.27\textheight]{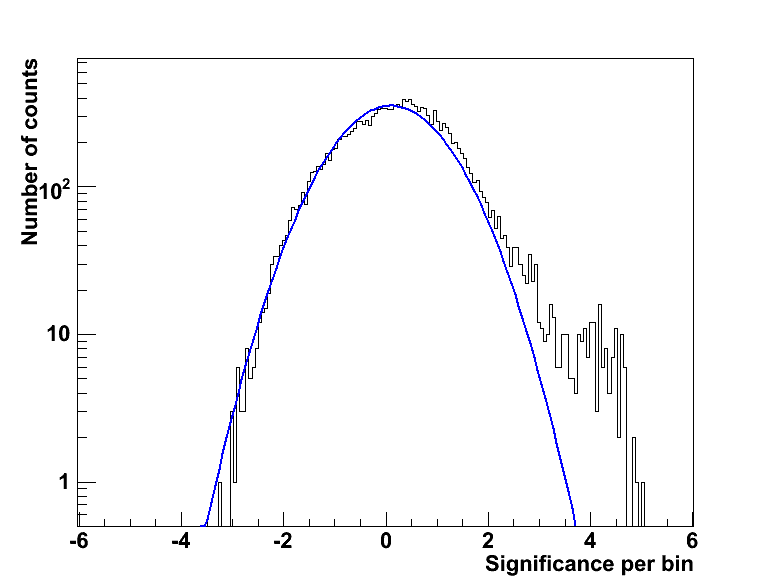}
\caption{Significance Distribution (Gaussian fit taken from the FOV excluding source: Mean=0.1, RMS=1)}
\end{figure}

 \begin{figure}
\includegraphics[height=.29\textheight]{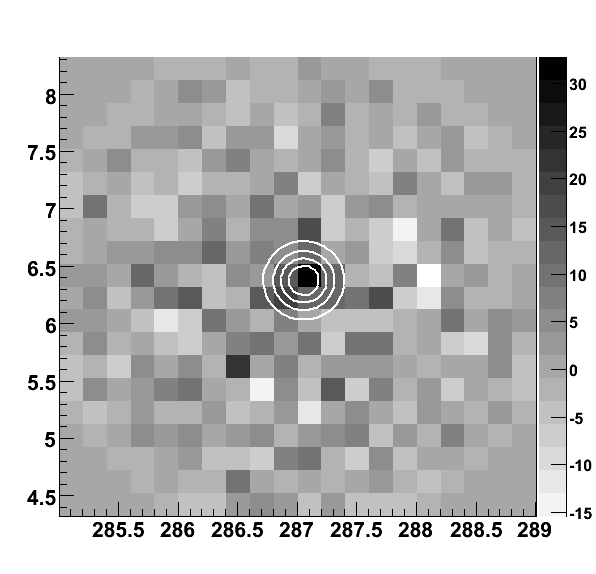}
\caption{2-Dimensional Gaussian fit to uncorrelated excess map (0.2$^\circ$ bins)}
\end{figure}

To investigate the position and extension of this source, a symmetrical two-dimensional
Gaussian function was fitted to the uncorrelated excess map for
this field. 
The best-fit position lies at RA=287.05$^\circ$ $\pm $0.05$^\circ$ (19$^h$ 08$^m$12$^s$) and 
DEC=6.37$^\circ$ $\pm $0.04$^\circ$ (06$^\circ$22$'$12$''$) with a 68\% excess containment radius (Point Spread Function not deconvolved) of $\sigma_{68\%}$= 0.19$^\circ$ $\pm $0.04$^\circ$ (the corresponding measurement for the Crab Nebula is $\sigma_{crab\_68\%}$ = 0.07$^\circ$), these measurements indicate that the source is not point-like and that the VERITAS best-fit position is fully compatible with previous results.

Figure 3 displays the uncorrelated excess plot of the field of view with 2D Gaussian contours overlaid. The source appears to be asymmetric therefore the 2-Dimensional Gaussian fit values should be taken as first approximations.
A more detailed study of source extension/morphology is ongoing.


\subsection{Counterpart Discussion}

\begin{figure}
\includegraphics[height=.32\textheight]{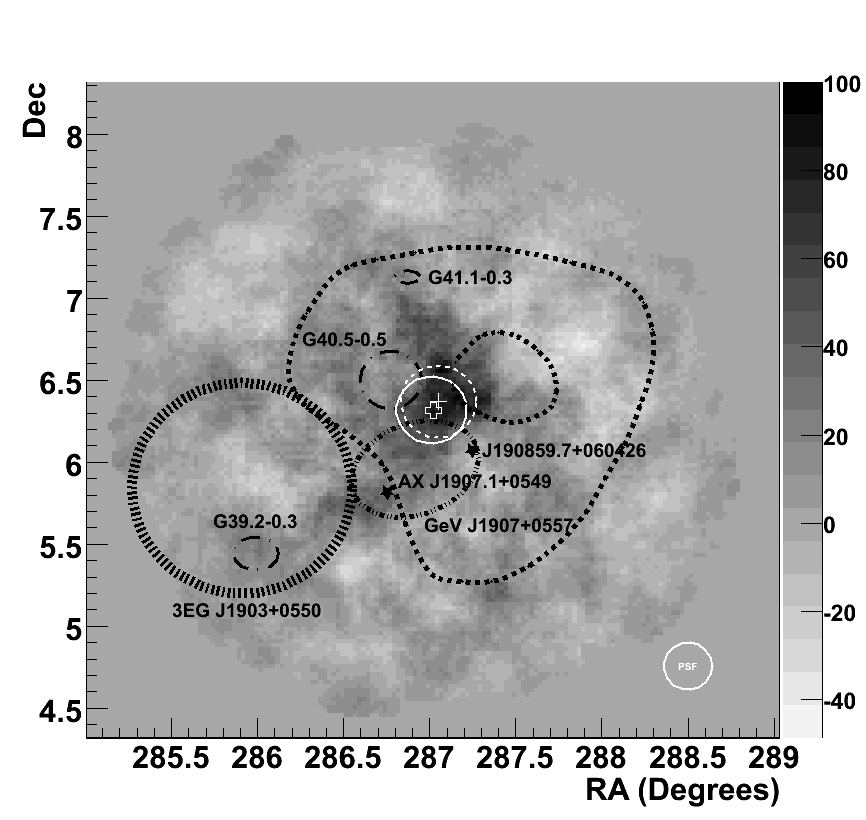}
\caption{Excess Map of FOV around MGRO J1908/HESS J1908. All SNR's are shown as black dot-dashed ellipses, the EGRET GeV J1908+0557 error ellipse is shown with a dot-dashed line, the 3EG J1903+0550 error circle is shown with a dashed line, the two X-Ray sources (AX J1907.1+0549 and 1RXS J 190859.7+060426) are overlaid as stars. The two ground-based arrays best-fit positions are shown, the H.E.S.S. quoted position (white open cross) along with the $\sigma_{src}$ excess containment radius (white solid circle) and the VERITAS best-fit position (white cross) with $\sigma_{68\%}$ excess containment radius (white dashed-circle). Finally the 5$\sigma$ and 7$\sigma$ Milagro contours are overlaid with thick dashed lines.}
\end{figure}

Figure 4 displays the VERITAS $\gamma$-ray excess map with potential multi-wavelength counterpart sources overlaid.
The Milagro 7$\sigma$ (outer) and 5$\sigma$ (inner) contours are shown overlain on the field along with the previously quoted H.E.S.S. position and extension value.
Potential TeV $\gamma$-ray sources located in this region include the Supernova Remnant (SNR) G40.5-0.5 \cite{green_snr} {which is located at a distance between 5.5-8.5 kpc, this remnant has been observed emitting non-thermal radio emission \cite{downes} and its level of extension somewhat overlaps with the VERITAS best-fit location. Two additional SNR are marked within the field-of-view but cannot be readily identifiable with the emission region.
The ASCA X-ray source AX J1907+0557 and the ROSAT source 1RXSJ190859.7+0602426 are overlaid, with Roberts et al. \cite{Roberts_1} 
showing AX J1907+0557 as an extended X-ray source compatible with the EGRET GeV source. 

\section{Conclusions}

VERITAS reports a detection of TeV gamma-rays from MGRO J1908+06/HESS J1908+063 at the
4.9$\sigma$ level confirming that this is a Very High Energy gamma-ray emitter. The
source location as measured by VERITAS is RA=287.05$^\circ$ (19$^h$ 08$^m$ 12$^s$) and DEC=6.37$^\circ$ (06$^\circ$22$'$12$''$) with
an extended emission region. Further analysis of this source is ongoing with
aims for flux/spectral measurements to be finalised for the possibility of future publication.


\begin{theacknowledgments}
This research is supported by grants
from the U.S. Department of Energy, the
U.S. National Science Foundation, and
the Smithsonian Institution, by NSERC in
Canada, by PPARC in the UK and by
Science Foundation Ireland.
\end{theacknowledgments}



\bibliographystyle{unsrt}

\bibliography{mgrohess-paper}


\end{document}